\documentclass[prb,aps,superscriptaddress,reprint]{revtex4-1}
\usepackage{graphicx}
\usepackage{amsmath}
\usepackage{amsbsy}
\usepackage{float}
\usepackage{amssymb}
\usepackage[makeroom]{cancel}
\usepackage[english]{babel}
\usepackage{color}
\usepackage{bm}
\usepackage{soul}

\begin{document}
\title{Durability of the superconducting gap\\ in Majorana nanowires under orbital effects of a magnetic field}

\author{P. W{\'o}jcik}
\affiliation{AGH University of Science and Technology, Faculty of Physics and Applied Computer Science, al. A. Mickiewicza 30, 30-059 Krakow, Poland}

\author{M. P. Nowak}
\affiliation{AGH University of Science and Technology, Academic Centre for Materials and Nanotechnology, al. A. Mickiewicza 30, 30-059 Krakow, Poland}

\date{\today}

\begin{abstract}
We analyze the superconducting gap in semiconductor/superconductor nanowires under orbital effects of a magnetic field in the weak- and 
strong-hybridization regime using a universal procedure which guarantees the stationarity of the system, i.e., vanishing of the supercurrent induced 
by a spatially varying vector potential. We perform minimization of the free energy with respect to the vector potential which allows for taking into 
account the orbital effects even for systems with intrinsically broken spatial symmetry. For the experimentally relevant scenario of a strongly 
coupled semiconductor/superconductor system, where the wave function of the charge carriers hybridizes between the two materials, we find that the 
gap closes due to the orbital effects in a sizable magnetic field in correspondence with the recent experiment [S. M. Albrecht, et al., Nature 531, 
206 (2016)].

\end{abstract}

\maketitle

\section{Introduction}
Over the last years Majorana bound states (MBSs), the simplest non-Abelian particles, have attracted the growing interest in the condensed matter 
physics due to their potential application in fault-tolerant topological quantum computation \cite{kitaev_fault-tolerant_2003,Alicea}. Among 
theoretical proposals of MBSs creation, including those based on topological insulator-superconductor junctions \cite{Fu_Kane}, graphene-like 
systems\cite{graphene_majorana,graphene_majorana2,graphene_majorana3,graphene_majorana4} or a chain of magnetic 
atoms\cite{NadjPerge,Pientka,Shiba,Klinovaja_atomic_chain}, the 
most promising one is related to semiconductor nanowires in which topological superconductivity can be induced by the proximity effect 
in the presence of both the spin-orbit interaction\cite{Wojcik_Goldoni} and the Zeeman effect\cite{oreg_helical_2010, sau_generic_2010, Lutchyn}.

Although the existence of MBSs localized at the ends of a nanowire has been confirmed in experiments \cite{mourik_signatures_2012, 
deng_anomalous_2012, finck_anomalous_2013, churchill_superconductor-nanowire_2013, 
chen_experimental_2017,albrecht_exponential_2016, zhang_quantized_2017}, a typical experimental setup with a single wire proximitized to a
superconductor is insufficient for topological quantum computation. Even an elementary braiding 
operation requires at least a three terminal junction. Only recently, the extensive progress in synthesis of semiconductor nanowires with a thin 
Aluminum shell \cite{krogstrup_epitaxy_2015, gazibegovic_epitaxy_2017} have directed experimental studies towards multiterminal devices 
\cite{plissard_formation_2013, fadaly_observation_2017, gazibegovic_epitaxy_2017} which can be used as prototypes for  topological quantum gates 
\cite{alicea_non-abelian_2011, heck_coulomb-assisted_2012, hyart_flux-controlled_2013}. The pristine interface between semiconductor 
and superconductor in these heterostructures\cite{kjaergaard_transparent_2017}, on the one hand guarantees the hard superconducting gap in the 
nanowire\cite{chang_hard_2015}, and on the other enables arbitrary alignment of the magnetic field without destroying  
superconductivity\cite{albrecht_exponential_2016}. In practice, the perpendicular orientation of the magnetic field is the most desirable for 
multiterminal structures as only this alignment allows for inducing the topological phase in all the nanowire branches simultaneously. In this case, 
the orbital effects of the magnetic field become of high importance since, as shown in Ref.~\onlinecite{dmytruk_suppression_2017}, even for the 
magnetic field aligned with the nanowire axis, they lead to the substantial modification of the topological phase diagram.

The standard way of theoretical treatment of the orbital effects is carried out by the incorporation of the canonical momentum with the appropriate 
vector potential into the Hamiltonian. However, recent years showed that the numerical adaptation of this method to the Hamiltonian with the 
particle-hole symmetry, even via the Peierls phase -- which makes it robust against discretization errors -- leads to ambiguous conclusions differing 
between subsequent studies. 

In the first theoretical analysis of the robustness of MBSs with respect to the magnetic field~\cite{lim_magnetic-field_2012} the authors stated 
that even a few degrees of magnetic field titling with respect to the nanowire axis destroys the zero energy modes due to the orbital effects. 
However, the following paper~\cite{osca_majorana_2015} indicated that correct discretization of the Bogoliubov-de Gennes Hamiltonian with the use of
the covariant derivative prevents the numerical artifacts which wrongly suggest that MBSs are easily destroyed by the orbital effects. As a result 
the topological phase can survive sizable vertical field tilting in favor of creation of the zero energy modes. The theoretical treatment of the 
orbital effects becomes cumbersome in more realistic models of heterostructures that aim at an accurate description of both the semiconducting 
nanowire and the metallic superconducting shell. Particularly, as argued by Nijholt and Akhmerov in Ref.~\onlinecite{nijholt_orbital_2016}, the 
orbital effects of a magnetic field in the heterostructures with a thin Al shell in the long-junction regime, break spatial and chiral symmetries of 
the Hamiltonian which leads to the tilting of the band structure and closing of the superconducting gap even for weak magnetic fields.

In the light of the ongoing debate, within this paper we provide careful analysis of the orbital effects on the closing of the superconducting gap. 
Our study settles the aforementioned ambiguity both for the systems in the weak-coupling regime and in the strong-coupling limit applicable to the 
recently studied nanowires~\cite{krogstrup_epitaxy_2015, albrecht_exponential_2016, gazibegovic_epitaxy_2017,Reeg2017}. We point out the importance 
of finding the stationary state of the considered hybrid system -- a configuration where  the supercurrent induced by the vector potential is zero. 
Our method is based on the minimization of the free energy with respect to the vector potential which avoids the necessity of a prior choice of the 
vector potential origin. As we show, such an approach is crucial, since the wrong adaptation of the vector potential into the Hamiltonian, can lead 
not 
only to different phase diagrams but also to erroneous conclusions that the topological gap closes in the range of parameters where it is still open.
First, the proposed method is demonstrated for the homogeneous nanowires with an uniform energy gap that corresponds to the 
hybrid structures in the weak-coupling limit. Then, it is used to study the orbital effects in the realistic experimental setup with a thin Al shell 
where we take into  account the strong variance of the material parameters in the heterostructure, providing good agreement with the recent 
observations~\cite{albrecht_exponential_2016}. 

\section{Model}
We consider a two-dimensional (2D) semiconductor nanowire with the Rashba spin-orbit interaction and the 
superconducting pairing induced by the proximity to a thin superconducting layer. Assuming translational invariance in 
the $x$-direction, the Hamiltonian of the system in the basis $( \psi^{e \uparrow},  \psi^{h \downarrow}, \psi^{e \downarrow}, - \psi^{h \uparrow})$  is given by
\begin{eqnarray}
\hat{H} &=& \left ( \mathbf{\hat{p}} \frac{1}{m^*(y)} \mathbf{\hat{p}} - \mu (y)\right ) \sigma _0 \tau _z + \hat{H}_{SOI} \tau _z  \nonumber \\ 
&+&  \frac{1}{2}g(y) \mu _B B 
\sigma _z \tau_0 + \sigma_0 \pmb{\Delta}, 
\label{ham}
\end{eqnarray}
with,
\begin{equation}
\pmb{\Delta} = \left(\begin{array}{cc} 0 & \Delta(y)\\ \Delta^*(y) & 0 \end{array}\right),
\end{equation}
and where $m^*(y)$ is the spatially dependent effective mass, $g(y)$ is the g-factor, $\Delta(y)$ is the superconducting gap, $\mu(y)$ is the chemical 
potential, $B$ is the external magnetic field oriented in the $z$-direction, perpendicular to the nanowire plane, and $\sigma _i$, $\tau _i$ with 
$i=x,y,z$ are the Pauli matrices acting on spin- and particle-hole degrees of freedom, respectively. In Eq. (\ref{ham}), $\hat{H}_{SOI}$ is the Hamiltonian 
of the spin-obit interaction
\begin{equation}
 \hat{H}_{SOI}=\frac{1}{2} \left \{  \alpha(y), \bm{\sigma} \times \mathbf{\hat{p}} \right \},
\end{equation}
taken in the form that ensures hermiticity for the spatially varying strength of the coupling $\alpha(y)$, where $\bm{\sigma}=(\sigma 
_x, \sigma _y, \sigma _z)$ and $\{ \cdots \}$ denotes the anticommutator. 

The orbital effects of the magnetic field are included through the canonical momentum, $\mathbf{\hat{p}}=-i \hbar \nabla _{2D} + e \mathbf{A} \tau 
_z$ with the vector potential in the Lorentz gauge $\mathbf{A}=(-(y-y_0)B,0,0)$, where $y_0$ is the offset. Determination of $y_0$ will be discussed
in the further part of this article.
\begin{figure}[h!]
\center
\includegraphics[width = 8.5 cm]{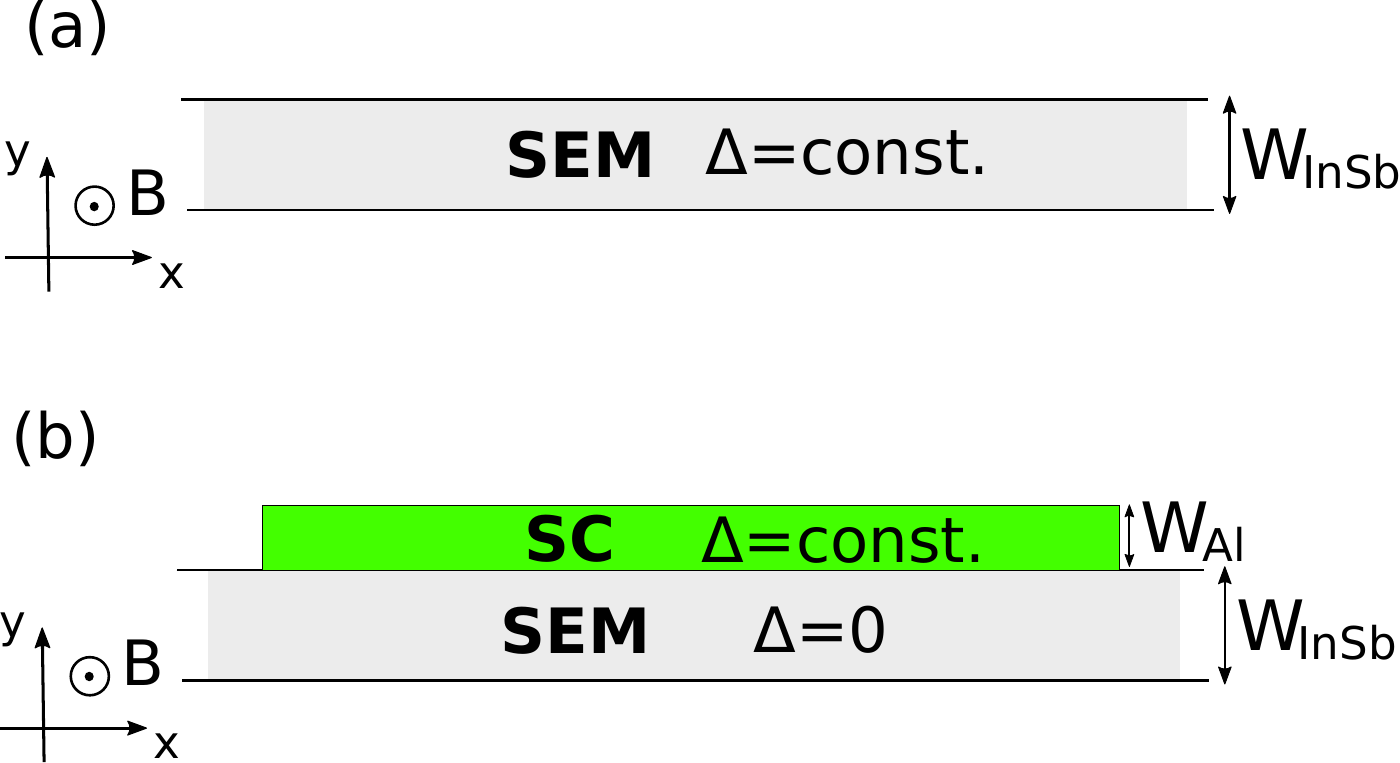}
\caption{
Schematic illustration of the considered systems: (a) homogeneous nanowire with the uniform energy gap and the width $W_{\text{InSb}}$ and (b) more 
realistic model of the semiconductor nanowire with the superconducting thin Al shell with the thickness $W_{\text{Al}}$. }
\label{fig1}
\end{figure}

In the following, we first consider a homogeneous nanowire with a uniform superconducting gap $\Delta$ inside the wire as presented in 
Fig.~\ref{fig1}(a). It corresponds to the weak-coupling regime\cite{sticlet_robustness_2017} with a non-transparent semiconductor/superconductor 
interface, where the superconductor can be considered as a perturbation providing solely the electron-hole coupling ($\Delta$) in the normal region. Then, we 
proceed to a more realistic model of the Majorana nanowire with a superconducting thin Al shell [see Fig.~\ref{fig1}(b)]. In this 
case all the $y$-dependent quantities in the Hamiltonian (\ref{ham}) have a form of step-like functions with values different for the two 
materials. For InSb nanowire we assume $m^*_{\text{InSb}}=0.014m_0$, $g_{\text{InSb}}=-51$, $\Delta_{\text{InSb}}=0$, 
$\alpha_{\text{InSb}}=50$~meVnm while the Al shell is characterized by $m^*_{\text{Al}}=m_0$, $g_{\text{Al}}=2$, 
$\Delta_{\text{Al}}=0.25$~meV, $\alpha_{\text{Al}}=0$.

Numerical diagonalization of the Hamiltonian (\ref{ham}) for the heterostructure from Fig.~\ref{fig1}(b) requires sophisticated discretization, due 
to the large difference in the effective masses and Fermi energies between semiconductor and superconductor. In order to minimize the numerical 
errors, the discretization is carried out on a non-uniform grid with two different lattice constants $a_{\text{InSb}}$ 
and $a_{\text{Al}}$ corresponding to the semiconducting and superconducting components, respectively. Then, the finite differences of the first and 
second derivatives along the $y$-axis are given by
\begin{eqnarray}
 &&\frac{\partial \psi}{\partial y} = \frac{\psi _{n+1}- \psi _{n-1}}{y_{n+1} - y_{n-1}}, \\
 && \frac{\partial}{\partial y} \left ( \frac{1}{m(y)}\frac{\partial \psi}{\partial y} \right ) = \frac{1}{2(y_{n+1}-y_{n-1})} \times  \nonumber 
\\
&& \bigg ( \frac{1}{m_{n+1/2}} \frac{\psi _{n+1} - \psi _n}{y_{n+1}-y_n} - \frac{1}{m_{n-1/2}} \frac{\psi _{n} - 
\psi _{n-1}}{y_{n}-y_{n-1}}
 \bigg ),
\end{eqnarray}
where $m_{n\pm1/2}$ denotes the average value of the effective mass between the grid points $n$ and $n\pm1$, respectively.

The numerical calculations are carried out on the rectangular grid with the following parameters: $a_{\text{InSb}}=2$~nm, 
$a_{\text{Al}}=0.01$~nm, $W_{\text{InSb}}=100$~nm and $W_{\text{Al}}=10$~nm unless stated otherwise. The vector potential is introduced 
into the numerical model through the 
Peierls substitution $t_{n,m} \rightarrow t_{n,m}\exp{(-i\frac{e}{\hbar} \int \mathbf {A} d \mathbf{l})}$. The numerical calculations for the 
homogeneous system were performed using the Kwant package~\cite{groth_kwant:_2014}.

\section{Results and discussion}
\subsection{Translation symmetry on a square lattice with a magnetic field}\
\label{sec:res1}
The discrete form of Hamiltonian (\ref{ham}) on the square lattice $(x_n,y_m)=(n,m)$ is given by
\begin{eqnarray}
 \hat{H} &=& \sum _{n} \left [ (4 t - \mu) \sigma _0 \tau _z + \frac{1}{2}g \mu _B B 
\sigma _z \tau_0 + \sigma_0 \pmb{\Delta} \right ] \hat{a}^{\dagger}_{n,n}\hat{a}_{n,n} \nonumber \\
&-& t \sigma_0 \tau _x \sum_{n,m} ( 
e^{i\theta^x_{n,m}}\hat{a}^{\dagger}_{n+1,m}\hat{a}_{n,m}+e^{i\theta^y_{n,m}}\hat{a}^{\dagger}_{n,m+1}\hat{a}_{n,m}+h.c ) \nonumber \\
&+& t_{SO}  \sum_{n,m} ( e^{i\theta^y_{n,m}} \sigma _x \tau _z \hat{a}^{\dagger}_{n,m+1}\hat{a}_{n,m} + h.c )\nonumber \\ 
&-& t_{SO}  \sum_{n,m} ( e^{i\theta^x_{n,m}} \sigma _y \tau _z \hat{a}^{\dagger}_{n+1,m}\hat{a}_{n,m} + h.c ), 
\label{ham:tba}
\end{eqnarray}
where $\hat{a}^{\dagger}_{n,m}$ and $\hat{a}_{n,m}$ are the creation and annihilation operators on site $(n,m)$, $t=1/2m^*a^2$, 
$t_{SO}=-i\alpha/2a$ with $a$ being the lattice constant and $\theta^{x(y)} _{n,m}=-eA^{x(y)}_{n,m}/\hbar$ is the Peierls phase along the $x(y)$-axis 
in the magnetic field $\mathbf{B}=\nabla \times \mathbf{A}$.

Note that the Hamiltonian (\ref{ham:tba}) is no longer invariant under the translation by one unit lattice vector because the corresponding vector 
potential $A_{n,m}$ is not invariant under this discrete translation even though the magnetic field $\mathbf{B}$ itself might be. It can be easily 
verified that the translation operators 
\begin{eqnarray}
 \hat{T} _x &=& \sum _{n,m}\hat{a}^{\dagger}_{n+1,m}\hat{a}_{n,m}e^{i\theta^x_{n,m}}\sigma _0 \tau _z, \\
 \hat{T} _y &=& \sum _{n,m}\hat{a}^{\dagger}_{n,m+1}\hat{a}_{n,m}e^{i\theta^y_{n,m}}\sigma _0 \tau _z,
\end{eqnarray}
do not commute with the Hamiltonian (\ref{ham:tba}), $[\hat{T}_{x(y)},\hat{H}]\neq0$.
To recover the translational invariance, the new \textit{magnetic translational operators}\cite{Topological} have to be constructed with the general 
form given by
\begin{eqnarray}
 \hat{T}^M _x &=& \sum _{n,m}\hat{a}^{\dagger}_{n+1,m}\hat{a}_{n,m}e^{i\chi^x_{n,m}} \sigma _0 \tau _z, \\
 \hat{T}^M _y &=& \sum _{n,m}\hat{a}^{\dagger}_{n,m+1}\hat{a}_{n,m}e^{i\chi^y_{n,m}} \sigma _0 \tau _z. 
\end{eqnarray}
The phases $\chi^{x(y)}_{n,m}$ are determined by the requirement $[\hat{T}^M_{x(y)},\hat{H}]=0$ which leads to
\begin{eqnarray}
 \chi^x_{n,m} = \theta^x _{n,m} + m\phi _{n,m}, \:\:\:\:
 \chi^y_{n,m} = \theta^y _{n,m} - n\phi_{n,m},
\end{eqnarray}
with 
\begin{equation}
\phi _{n,m}=\frac{e}{\hbar} (\theta ^x_{n,m}+\theta ^y_{n+1,m}-\theta ^x_{n,m+1}-\theta ^y_{n,m}),
\end{equation}
being the magnetic flux per unit cell. For the considered Lorentz gauge $\mathbf{A}_{n,m}=(-maB,0,0)$ the magnetic translational operators are
\begin{eqnarray}
 \hat{T}^M _x &=& \sum _{n,m}\hat{a}^{\dagger}_{n+1,m}\hat{a}_{n,m} \sigma _0 \tau _z, \\
 \hat{T}^M _y &=& \sum _{n,m}\hat{a}^{\dagger}_{n,m+1}\hat{a}_{n,m}e^{in\phi} \sigma _0 \tau _z,
 \label{tmy}
\end{eqnarray}
where $\phi _{n,m}=\phi = eBa^2/\hbar$.\\
The operators $\hat{T}^M _x$, $\hat{T}^M _y$ commute with the Hamiltonian (\ref{ham:tba}) by construction. 
Physically, they correspond to the transformation of the Hamiltonian (wave function) due to the translation by one unit lattice vector along the 
$x$ and $y$-axis, respectively. 

The zero-field form of $\hat{T}^M _x$ indicates that the translation along the $x$-axis by an arbitrary vector does not change the 
Hamiltonian (\ref{ham:tba}). However the translation along the $y$-axis by the vector being integer multiple of the lattice vector  
$\mathbf{y}_d=(0,qa)$ with $q\in \mathbb {Z}$ [see Eq.~(\ref{tmy})] leads to the acquisition of the phase $\Phi _n=iqn\phi$, opposite for 
electron and holes, which depends on the position of site $(m,n)$ on the lattice. In this case, to ensure the gauge-invariance, the 
Hamiltonian (\ref{ham:tba}) has to be transformed by the unitary operator  
\begin{equation}
 \hat{U}=\left (
 \begin{array}{cc}
  e^{i\Phi_n} & 0 \\
  0 & e^{-i\Phi_n}
 \end{array}
 \right ).
\end{equation}
The transformation $\hat{U} \hat{H} \hat{U}^\dagger$ modifies the superconducting pair potential to the form $\Delta \rightarrow \Delta e^{i2\Phi 
_n}$, which guarantees that the supercurrent $j_s=\frac{1}{2}\nabla \Phi -e\mathbf{A}$ does not change. 
Specifically, if $j_s=0$ this transformation preserves the stationarity of the system. As we will show in the next sections, the requirement of 
stationarity is indispensable and has to be incorporated in the simulations of the Majorana nanowires that include the orbital effects 
through the vector potential.

\subsection{Homogeneous nanowire}
We start from the simple homogeneous nanowire with the uniform energy gap as presented in Fig.~\ref{fig1}(a). The induced energy 
gap is taken to be $\Delta_{\text{InSb}}=0.25$~meV.

\subsubsection{Symmetric system}
\begin{figure}[h!]
\center
\includegraphics[width = 8.5 cm]{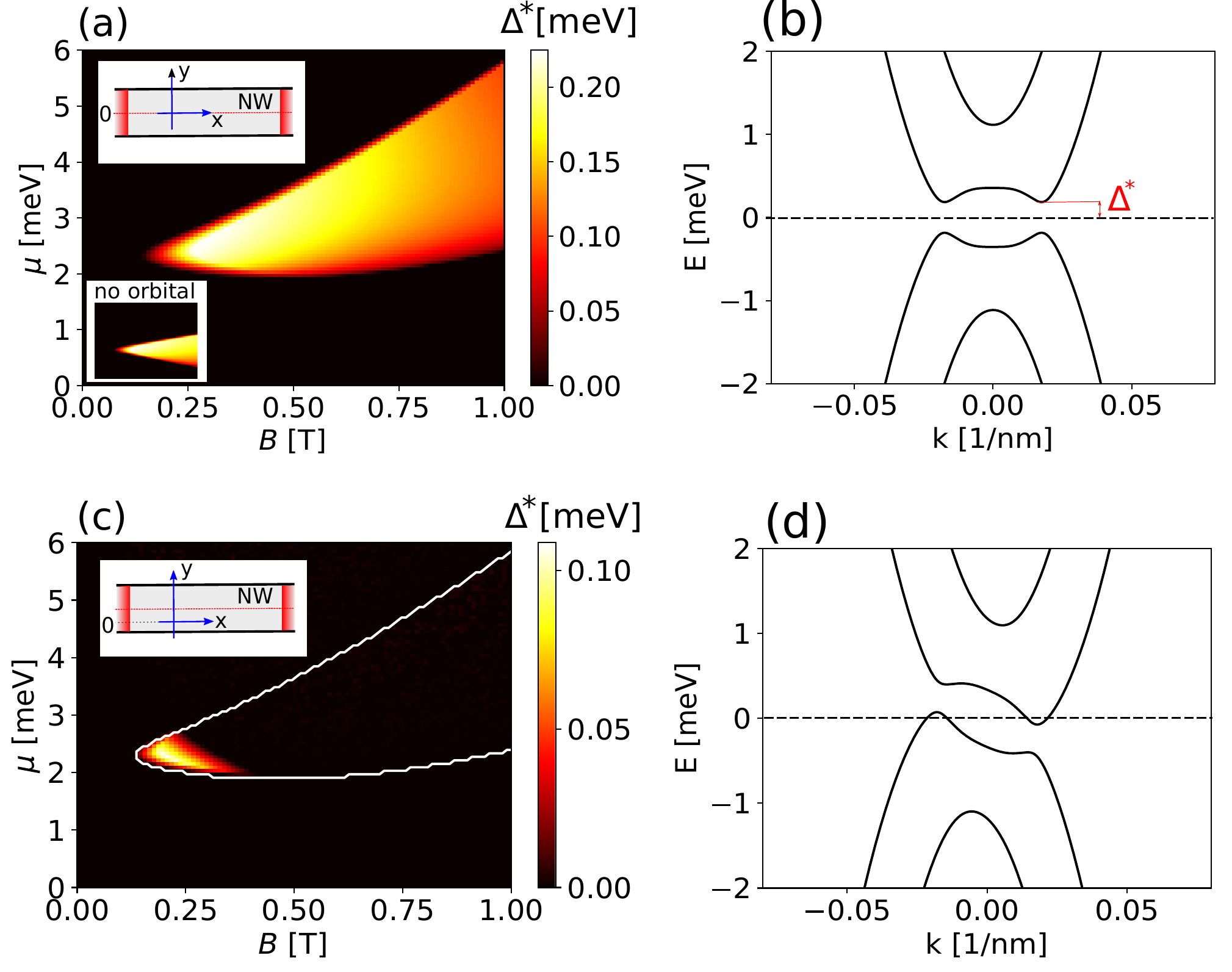}
\caption{
(a),(c) Topological energy gap $\Delta ^*$ as a function of the magnetic field $B$ and the chemical potential $\mu$. (b),(d) Dispersion relations 
$E(k)$ calculated for $\mu=2.3$~meV and $B=0.4$~T. Top panels: results for the nanowire localized symmetrically with respect to the $x$-axis and the 
vector potential offset $y_0=0$, bottom panels: results for the nanowire shifted by the vector $(0,y_{d})$ with $y_{d}=8$~nm and 
$y_0=0$ (see the insets in left-top corners). The inset in the left-bottom corner of panel (a) presents topological energy gap 
$\Delta ^*(B,\mu)$ calculated without orbital effects - the scale on the axes is the same as on the main panel (a).
}
\label{fig2}
\end{figure}

Let us first consider the nanowire localized symmetrically with respect to the $x$-axis and put the offset of the vector potential $y_0=0$ [see the 
top-left inset of Fig.~\ref{fig2}(a)]. 
The main quantity under consideration is the topological gap $\Delta ^*$ determined from the gapped Dirac cones at $k\neq 0$ in the topological 
phase as presented in Fig.~\ref{fig2}(b). In Fig.~\ref{fig2}(a) we plot the map of $\Delta^*$ calculated with the inclusion of the orbital effects as 
a function of the magnetic field and the chemical potential. The shape of the topological phase contour strongly deviates 
from the one obtained with sole Zeeman interaction [see the bottom-left inset to Fig.~\ref{fig2}(a)] which results from the renormalization of the 
effective mass, spin-orbit coupling and chemical potential due to the orbital effects\cite{Nowak_wojcik}. Dispersion relation for an exemplary set of 
parameters ensuring the topological phase is presented in Fig.~\ref{fig2}(b). 

Now, let us transpose the nanowire by the vector $\mathbf{y}_d=(0,y_{d})$ with $y_{d}=8$~nm keeping the vector potential offset $y_0=0$ [see the 
inset in Fig.~\ref{fig2}(c)]. Equivalently, we can leave the position of the nanowire unchanged and transform the vector 
potential changing the offset $y_0=-y_{d}$. In this case, if we do not take care of the appropriate transformation of the Hamiltonian, the 
particle and hole components acquire opposite phases from the magnetic field as described in \ref{sec:res1}. This leads to tilting of the band 
structure [see Fig.~\ref{fig2}(d)] which corresponds to generation of a suppercurrent 	\cite{Chirolli1,Chirolli2}, 
to a significant reduction of the parameter space where the topological gap is nonzero [c.f. 
Fig.~\ref{fig2}(c) with Fig.~\ref{fig2}(a)] and the striking conclusion that the magnetic field closes the superconducting gap which makes the 
creation of Majoranas impossible.

Note that the correct treatment of the Bogoliubov-de Gennes Hamiltonian under the gauge transformation $\mathbf{A} \rightarrow \mathbf{A} + 
\nabla \Lambda$ requires an appropriate transformation of the wave function $(\psi _e, \psi _h) \rightarrow ( \psi _e e^{-i\frac{e\Lambda}{\hbar}}, 
\psi _h e^{i\frac{e\Lambda}{\hbar}})$ and the superconducting gap $\Delta \rightarrow \Delta e^{-2 i\frac{e \Lambda}{\hbar}}$. As we checked, in the 
case of nanowire displacement by the vector $\mathbf{y}_{d}$ when $\Lambda=-y_{d} B x$, the numerical results do not depend on the 
choice of $y_{d}$ giving the topological phase diagram as presented in Fig.~\ref{fig2}(a) even when the system is located as in the inset 
to Fig.~\ref{fig2} (c). Seemingly, the assurance of the gauge invariance by the aforementioned transformation solves the problem of artificial gap 
closing. However, in practice the gauge invariance only ensures that the results 
stay unchanged under the transformation $\mathbf{A} \rightarrow \mathbf{A} + \nabla \Lambda$ but does not determine the primordial alignment of the 
system with respect to the vector potential. In other words, we could as well assume that the results from Fig.~\ref{fig2}(c)(d) present the physical 
solution that remains unchanged upon the gauge transformation.

As we present in Sec.~\ref{sec:res1}, the incorporation of the magnetic field should rather be associated with an additional condition for the 
vector potential which guarantees stationarity of the system. Physically, it can be achieved by zeroing of the supercurrent $j_s$ due to appropriate 
choice of the vector potential. Determination of $j_s$ for the considered system comes down to $j_s=\sum 
_n \frac{e}{i\hbar} \langle \psi_{n}|[x\sigma_0\tau_z,\hat{H}]|\psi_{n} \rangle$ and requires calculation of the eigenvectors. 
As the presence of the supercurrent corresponds rather to the excited than the ground state, the condition 
$j_s=0$ can be alternatively found in a simpler way by minimizing the free energy which for superconducting nanostructures takes the 
form\cite{Kosztin}
\begin{eqnarray}
 &&\mathcal{F}[\Delta(\mathbf{r}),\mathbf{A}(\mathbf{r})] = E_g + 2\sum_i E_if_i \nonumber \\
  &-& 2k_BT\sum_i[f_i \ln{f_i}+(1-f_i)\ln (1-f_i)] + \mathcal{F_B}  
 \label{free_energy}
\end{eqnarray}
where $E_i$ are eigenvalues of the Hamiltonian (\ref{ham}), $f_i=[\exp(E_i / k_B T)+1]^{-1}$ is the Fermi distribution function and $T$ is a 
temperature. $\mathcal{F}_B$ is a positive magnetic field exclusion energy due to the screening supercurrent induced by the magnetic field
\begin{equation}
 \mathcal{F}_B = \int \frac{(\mathbf{B}(\mathbf{r})-\mathbf{B}_a)^2}{8\pi} \mathbf{dr}.
\end{equation}
In practice, to avoid divergence in the expression (\ref{free_energy}), we calculate it with the respect to the free energy 
$\mathcal{F}_N$ of the corresponding normal state, $\delta \mathcal{F} = \mathcal{F}-\mathcal{F}_N$. Note, that for $T=0$, $\delta \mathcal{F}$ 
reduces to the formula for the condensation energy 
\begin{equation}
 \mathcal{E} _{\Delta} = - \sum _i (E_i - \xi _i) + E_{\Delta}
\end{equation}
where $\xi _i$ are the eigenvalues of the Hamiltonian (\ref{ham}) in the normal state while $E_{\Delta}$ can be treated as the energy reference 
level independent on $\mathbf{A}(\mathbf{r})$. 

\begin{figure}[h!]
\center
\includegraphics[width = 5.5 cm]{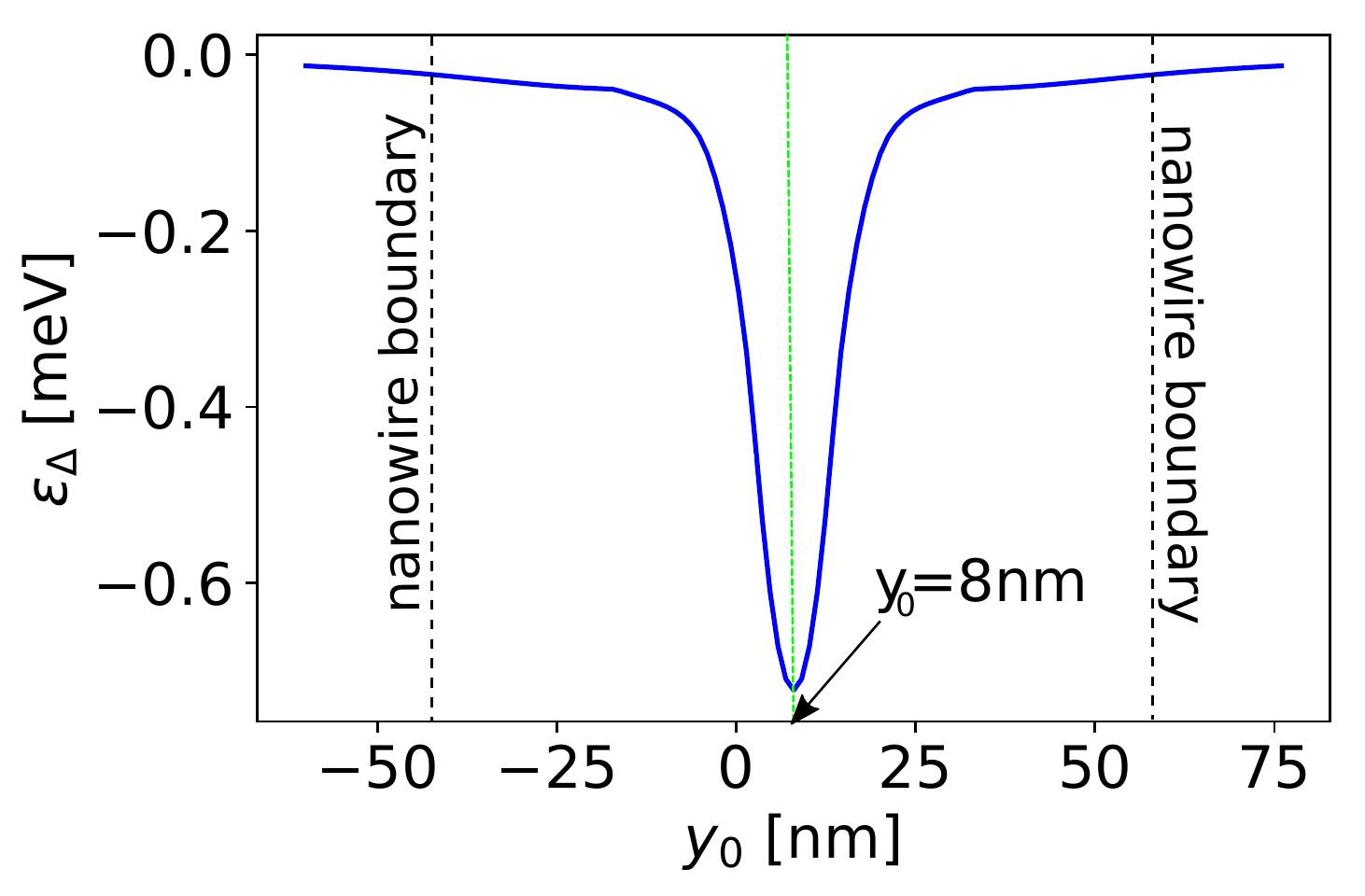}
\caption{
Condensation energy $\mathcal{E}_{\Delta}$ as a function of the vector potential offset $y_0$. The minimum of $\mathcal{E}_{\Delta}$  corresponding to 
the stationary state with $j_c=0$ is located in the middle of nanowire marked by the green vertical line.  Results for $\mu=2.3$~meV and $B=0.4$~T.
}
\label{fig3}
\end{figure}

Minimization of $\mathcal{E} _{\Delta}$ with respect to the vector potential is the key point for the inclusion of the orbital effects, which 
guarantees the stationarity. It can be done by appropriate choice of the vector potential offset $y_0$\footnote{As presented in 
sec.~\ref{sec:res1} only the translation along the $y$-axis induces the spatially dependent phase and generates the supercurrent}. Due to the 
reflection symmetry with respect to the $x$-axis, for the considered homogeneous nanowire with the chosen Lorentz gauge, the condition $j_s=0$ 
requires the 
vector potential offset $y_0$ being always positioned in the middle of the nanowire. This requirement is not 
met for the system presented in the inset of Fig.~\ref{fig2}(c) and the gap closing in this case is a result of the non-stationarity with $j_s \neq 
0$. In Fig.~\ref{fig3} we present the condensation energy $\mathcal{E} _{\Delta}$ as a function of the offset $y_0$ calculated for the nanowire 
shifted by the vector $\mathbf{y}_d$ [inset in Fig.~\ref{fig2}(c)]. Nanowire boundaries are depicted by the dashed lines. 
The distinct minimum of the condensation energy, which ensures stationarity of the system ($j_s=0$), is localized exactly 
in the middle of the nanowire as we previously inferred from the symmetry analysis. We also checked, that for the considered homogeneous nanowire, the 
position of this minimum does not change regardless of the magnetic field and the chemical potential values. The map of the topological gap 
determined by this method is exactly the same as presented in Fig.~\ref{fig2}(a), in which the stationarity is preserved by construction, and does 
not dependent on the translation vector $\mathbf{y}_d$.

\subsubsection{Broken symmetry}
For a symmetric system discussed above, even without employing the minimization procedure it was possible to guess the alignment of the vector 
potential that minimizes the supercurrent. This is however not possible when the spatial symmetry is intrinsically broken by, e.g., potentials in Eq. 
(\ref{ham}) such the charge distribution in the wire is not know a priori. To demonstrate that let us assume that the system is localized 
symmetrically about the $x$-axis with the vector potential offset $y_0=0$ and the symmetry is broken by the applied electric field $F_y$ oriented 
along the $y$-axis. 
\begin{figure}[h!]
\center
\includegraphics[width = 8.5 cm]{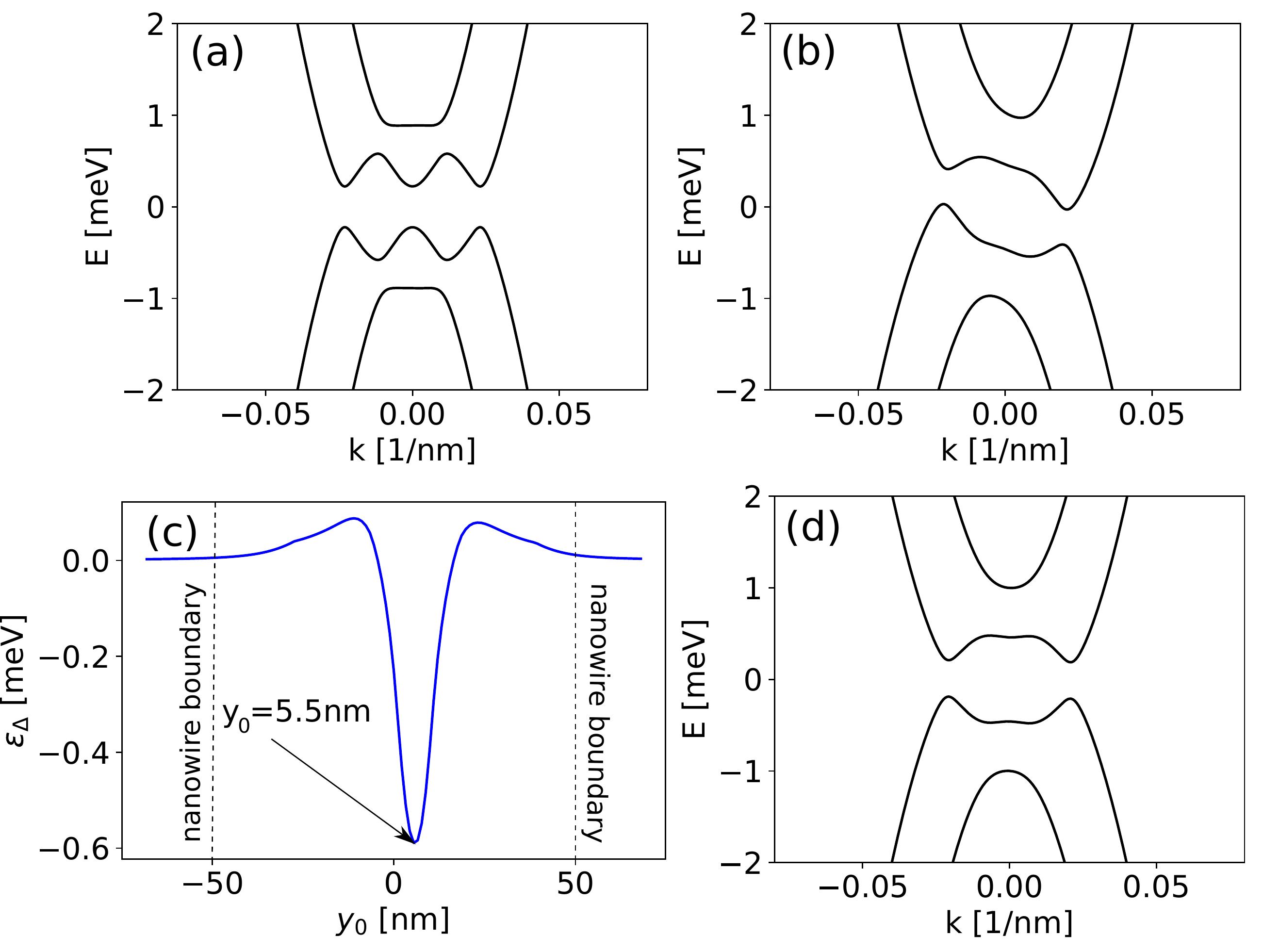}
\caption{
Band structures of the homogeneous nanowire [Fig.~\ref{fig1}(a)] in the presence of the transversal electric field $F_y$, calculated 
(a) without the orbital effects, (b) with the orbital effects and the vector potential offset $y_0=0$ and (d) with the orbital effects and the 
vector potential offset $y_0=5.5$~nm determined by the minimization of $\mathcal{E}_{\Delta}$ presented in (c). Results for $\mu=2.3$~meV, 
$B=0.4$~T and $F_y=0.6$~kV/cm.
}
\label{fig4}
\end{figure}

In this case, the sole presence of the transversal electric field without the magnetic orbital effects does not tilt the band structures -- $E(k)$
are fully symmetric with respect to $k=0$ as presented in Fig.~\ref{fig4}(a). The inclusion of the orbital effects breaks the chiral symmetry 
$\mathcal{C}=\tau _y\mathcal{R}_y$ with $\mathcal{R}_y=\sigma _y \delta(y+y')$ leading to the band tilting as presented  in Fig.~\ref{fig4}(b). 
Again, not taking care of the stationarity leads to the wrong conclusion that the gap closes already for low 
magnetic fields [see Fig.~\ref{fig4}(b)]. In fact, Fig.~\ref{fig4}(b) corresponds rather to the excited state with $j_c\ne0$. The full 
minimization of the condensation energy [see Fig.~\ref{fig4}(c)] clearly indicates that there is a vector potential offset $y_0=5.5$~nm that 
corresponds to the ground state and gives the band structure presented in Fig.~\ref{fig4}(d). In contrary to Fig.~\ref{fig4}(b) 
it does not display the gap closing.

\subsection{Semiconductor/supercondcutor heterostructure}
Finally, we turn our attention to the more realistic model presented in Fig.~\ref{fig1}(b), which explicitly treats the thin superconducting Al 
shell. Before discussing the orbital effects, we start from the brief overview of an appropriate semiconductor/superconductor interface 
parametrization which ensures the induced gap values as observed in recent experiments on nanowires with an epitaxial Al shell. Very recently the 
hybridization at the semiconductor/superconductor interface in Majorana devices has been studied by the self-consistent Schr{\"o}dinger-Poisson 
approach~\cite{Mikkelsen,Antipov}.
It has been pointed out that the quantitative description of the interface requires consideration of the band offset $V_{b}$ which results from the 
difference between the electron affinity of the semiconductor and the work function of the metal. As reported in the recent ARPES 
studies\cite{Antipov}, $V_{b}$ is negative for the epitaxially grown InAs/Al heterostructure supporting the scenario of the band bending which 
localizes the charge near the semiconductor/superconductor interface. This effect combined with the back-gate  
electric field substantially strengthens the hybridization between the states at the interface which otherwise is unfavorable due to the large 
difference between the effective masses and the chemical potentials in both materials. 
\begin{figure}[h!]
\center
\includegraphics[width = 6.5 cm]{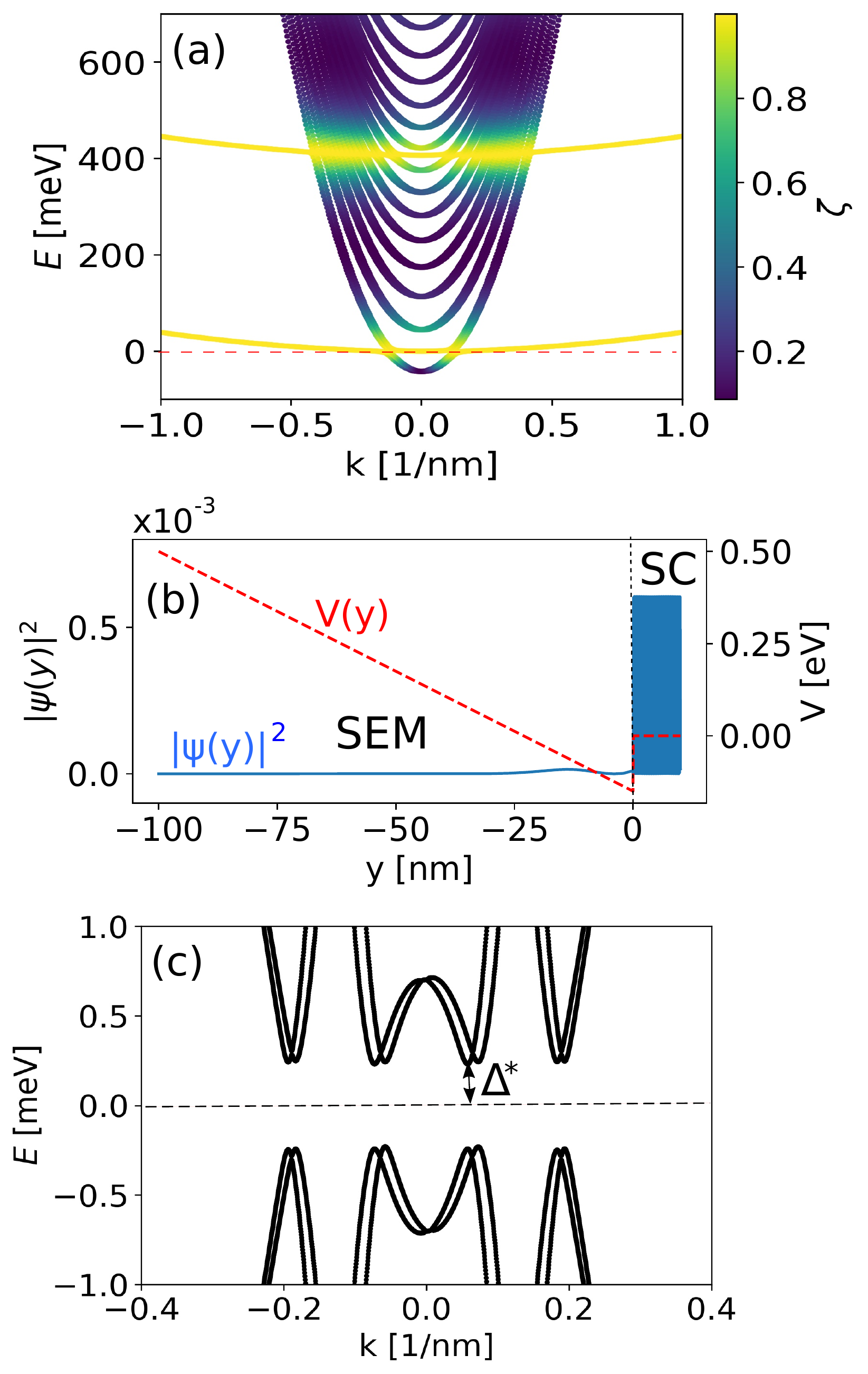}
\caption{
(a) Band structures of the InSb/Al nanowire. The color of the curves determines the amount of the wave function localized in the 
supercondcutor. (b) The potential profile $V(y)$ (red-dashed line) for the gate voltage $V_g=-0.5$~eV and the band offset $V_{b}=-0.15$~eV together 
with the squared value of the wave function calculated at $k _F \approx 0.1$~nm$^{-1}$ where the strong  hybridization is observed. (c) Band 
structure of the system in the superconducting state. Due to the strong hybridization the induced gap $\Delta ^*=0.25$~meV corresponds to the one 
assumed in the Al shell. Results for $B=0.4$~T without the orbital effects.
}
\label{fig5}
\end{figure}
However, even the accumulation of the charge near the semiconductor/superconductor interface does not guarantee the superconducting gap 
in the semiconductor. In fact, it is possible only if a strongly hybridized band crosses the Fermi level. This can be obtained by an 
appropriate adjustment of Al electronic states so that the energy of one of them crosses the Fermi level at the $k$ vector near the crossing point for the InSb 
lowest subband. This makes the induced gap sensitive to the Al layer thickness and the value of $V_{b}$~\cite{Mikkelsen}. In order to study the 
orbital effects in an experimentally observed gap regime we assume that the whole heterostructure is attached to the gate from the side of 
semiconductor, 
as in the experiment. For simplicity, we do not consider the Hartee potential - the full self-consistent calculations~\cite{Mikkelsen,Antipov} are 
time consuming and do not change the main conclusions of our study. Under these assumptions the negative gate voltage $V_g$ generates the 
triangular-shaped potential [see Fig.~\ref{fig5}(b), red dashed line] which together with negative band offset $V_{b}$ mimics the charge 
accumulation layer near the interface. In our calculations we take $V_g=-0.5$~V, $V_{b}=-0.15$~eV, and $\mu_{\text{InSb}}=0$. 
\begin{figure}[h!]
\center
\includegraphics[width = 5.5 cm]{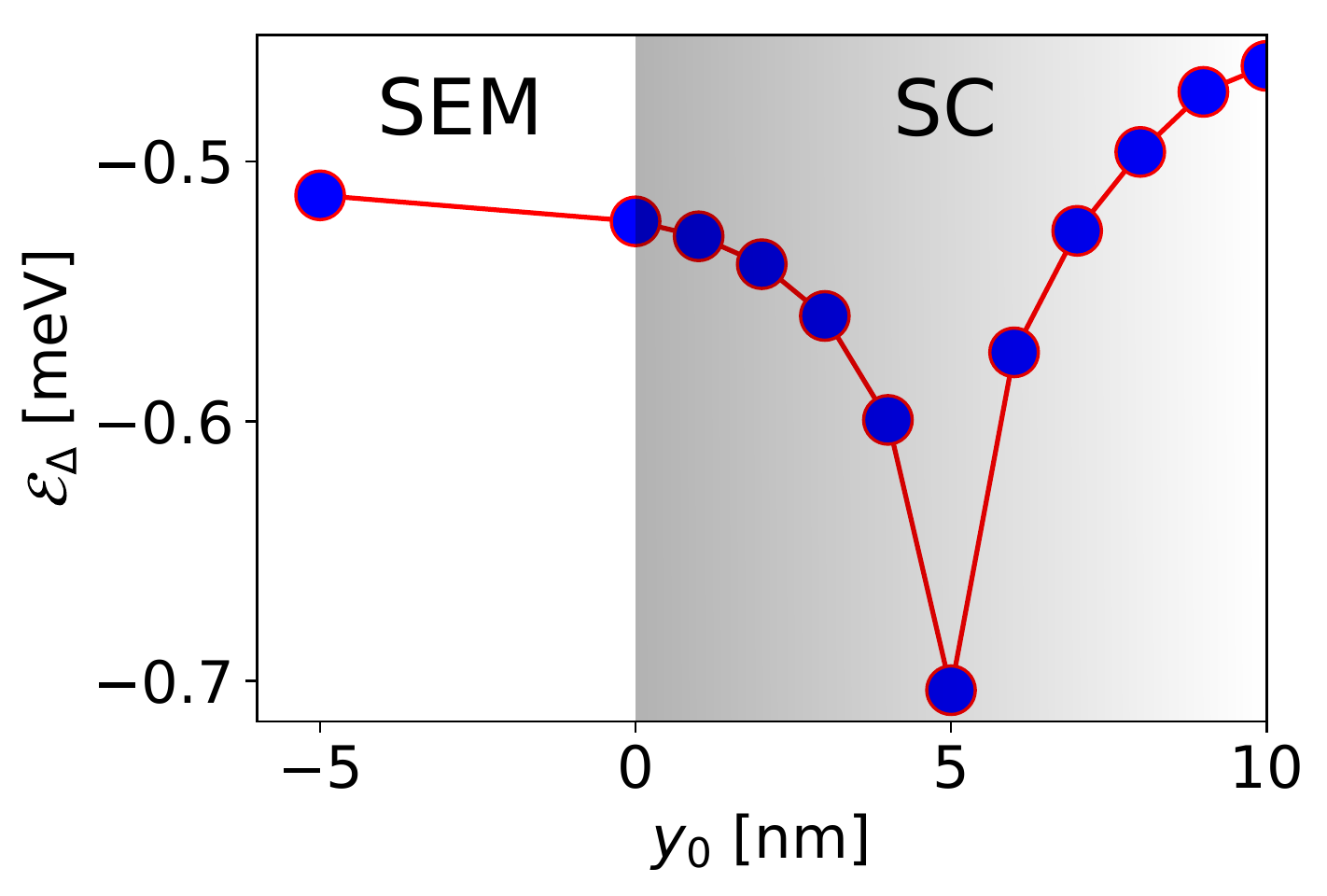}
\caption{
Condensation energy $\mathcal{E}_{\Delta}$ as a function of the vector potential offset $y_0$ for InSb/Al heterostructure. The superconductor is 
marked by the gray area. Results for $B=0.4$~T.
}
\label{fig6}
\end{figure}

\begin{figure}[th!]
\center
\includegraphics[width = 9. cm]{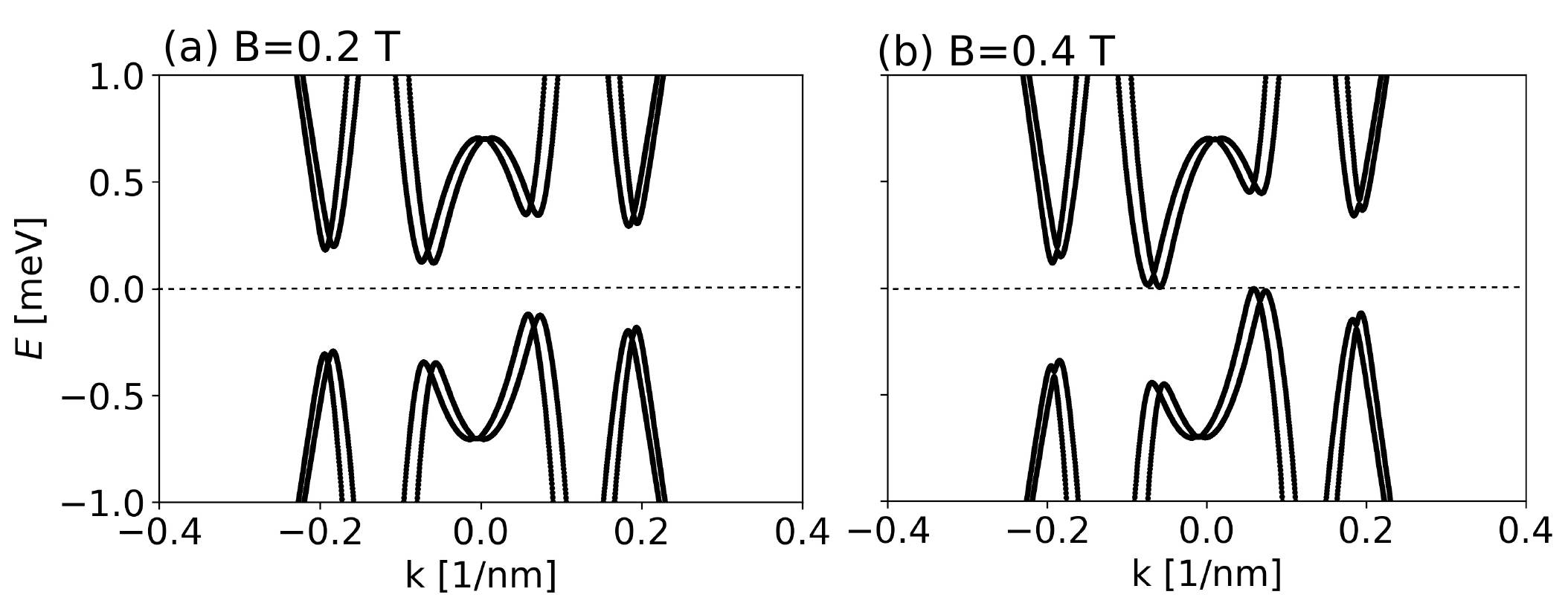}
\caption{
Band structures of the InSb/Al nanowire calculated with the inclusion of the orbital effects for (a) $B=0.2$~T and (b) $B=0.4$~T. 
}
\label{fig7}
\end{figure}

In Fig.~\ref{fig5}(a) we present the electron band structure of the InSb/Al nanowire obtained by the numerical solution of the Schr{\"o}dinger 
equation with the spatially dependent effective mass, chemical potential, spin-orbit constant and $g$-factor. The color of the curves determines how 
strongly different bands are coupled to the superconductor which is quantified by the amount of the wave function localized in 
superconductor 
\begin{equation}
 \zeta=\int _{0}^{W_{Al}} |\psi_n(y,k)|^2 dy.
\end{equation} 
The electronic state configuration which ensures strong hybridization at the Fermi level is obtained by the slight modification of the 
chemical potential to $\mu_{\text{Al}}=10.5$~eV from the bulk Al value $11.7$~eV\cite{Ashcroft}. This leads to the situation where the substantial 
part of the wave function is localized in the superconductor - see Fig.~\ref{fig5}(b), blue line. Then, as presented in Fig.~\ref{fig5}(c), the 
induced gap $\Delta ^*=0.25$~meV is close to that assumed for the Al shell.

The correct parametrization of the semiconductor/superconductor heterostructure which ensures the strong-coupling regime is crucial for understanding 
the impact of the orbital effects on the topological gap, compatible with the recent experimental observation. To show that  we start from 
calculations of the condensation energy $\mathcal{E}_{\Delta}$ as a function of the vector potential offset $y_0$ - see 
Fig.~\ref{fig6}. The dependence $\mathcal{E}_{\Delta}(y_0)$ exhibits the distinct minimum exactly at the middle of superconductor ($y_0=5$~nm)
Its position is independent on the magnitude of the magnetic field.
\begin{figure}[th!]
\center
\includegraphics[width = 5.5 cm]{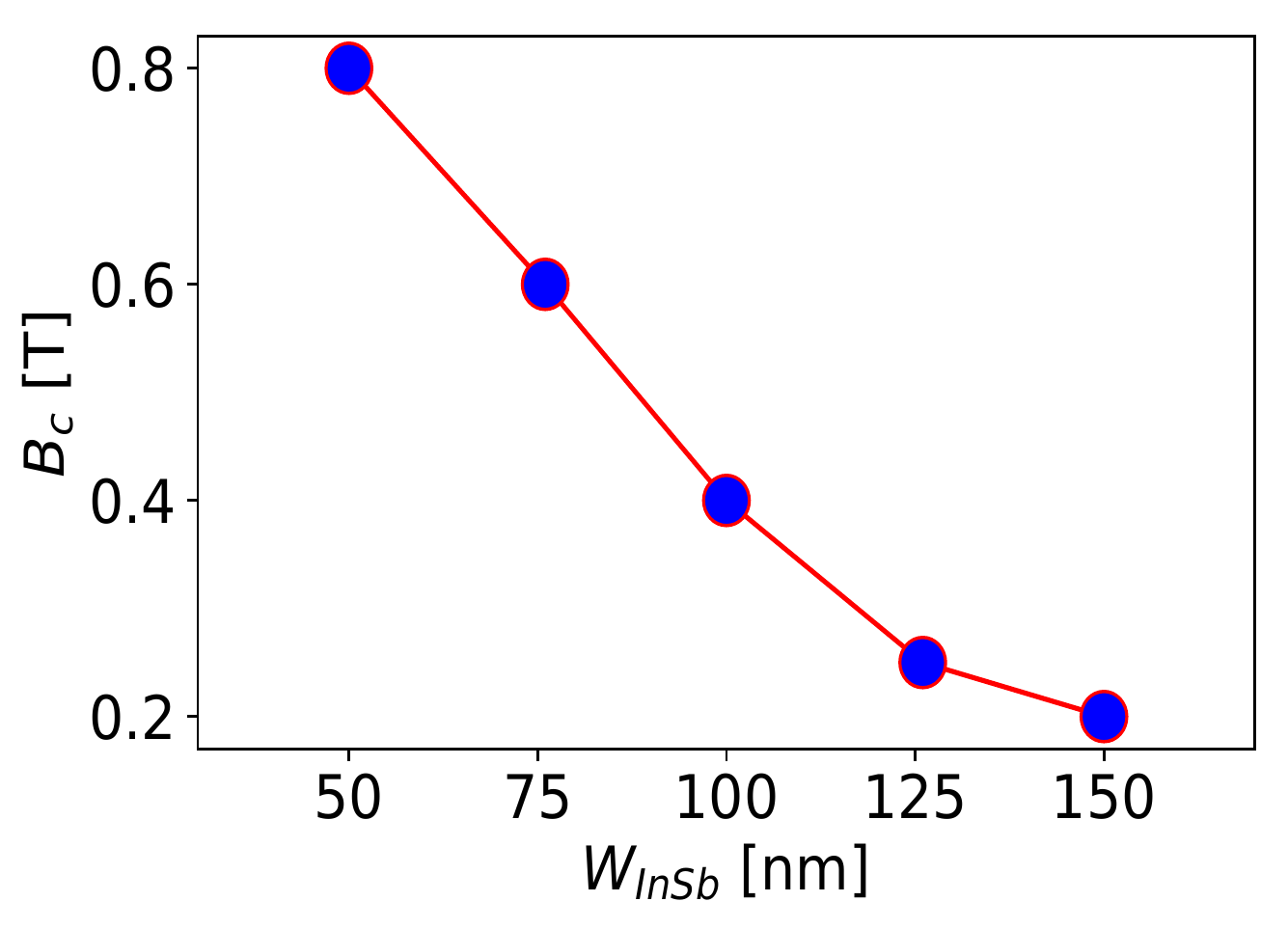}
\caption{
Critical field $B_c$ as a function of nanowire thickness $W_{\text{InSb}}$
}
\label{fig8}
\end{figure} 

Distinct hybridization of states in the considered strong-coupling regime localizes the wave function in the superconductor near the zero of the 
vector potential making the induced superconducting gap robust against the orbital effects. The band structures calculated with the inclusion of the 
orbital effects for different magnetic field magnitudes (see Fig.~\ref{fig7}) show that the gap closes for $B_c=0.4$~T close to the value reported in 
the experiment -- compare with Fig.~3(d) from Ref.~[\onlinecite{albrecht_exponential_2016}].
Note that, for the case of sole Zeeman interaction the physical restriction for the presence of the induced gap is the critical field of the Al shell 
above which superconductivity in Al is destroyed. Assuming that the superconducting properties of 10 nm thick Al shell do not strongly deviate from 
those for the bulk, we can estimate the Pauli paramagnetic limit based on the Clogston-Chandrasekhar formula $B_{c,P}=\Delta/\sqrt{2}\mu_B$, where 
$\mu_B$ is the Bohr magneton. For the assumed $\Delta=0.25$~meV, the estimated value $B_{c,P}=3$~T is significantly higher than the one observed 
experimentally.

Finally, in Fig.~\ref{fig8} we present the critical field as a function of the nanowire thickness $W_{\text{InSb}}$. As expected, the 
critical field increases for narrower wires. Nevertheless, even for small $W_{InSb}$ $B_c$ is much less than the paramagnetic limit 
$B_{c,P}$ indicating the significance of the inclusion of the orbital effects in reliable modeling of topological properties of hybrid nanowires.

\section{Summary}
We have analyzed the impact of the magnetic orbital effects on the superconducting gap closing in hybrid Majorana nanowires. We have demonstrated 
that the vanishing of the supercurrent -- stationarity -- is the necessary condition for a proper description of the semiconductor/superconductor 
hybrid under the magnetic field. We have proposed that the stationarity can be acquired by minimizing the free energy of the structure with respect 
to the vector potential. Following that procedure we have studied the superconducting gap in semiconductor/superconductor heterostructures in weak- 
and strong-coupling regimes. The proposed scheme avoids the need of an a priori choose of the location of the vector potential origin that might lead 
to erroneous conclusions or even can be simply impossible to determine for systems with intrinsically broken symmetry.  Finally, for the realistic 
heterostructure with a thin Al shell with the account taken to the strong variance of the material parameters, we have found that the critical field 
is comparable to that reported in the recent measurement for gated nanowires\cite{albrecht_exponential_2016}.

\begin{acknowledgements}
The authors acknowledge helpful discussions with Micha{\l} Zegrodnik.
P.W. was supported by National Science Centre, Poland (NCN) according to decision 2017/26/D/ST3/00109.
M.P.N. was supported by National Science Centre, Poland (NCN) according to decision DEC-2016/23/D/ST3/00394.
The calculations were performed on PL-Grid Infrastructure.
\end{acknowledgements}

%

\end{document}